# Hot Bubbles from Active Galactic Nuclei as a Heat Source in Cooling Flow Clusters

Marcus Brüggen[1] and Christian R. Kaiser[2]
[1] International University Bremen, Campus Ring 1, 28759 Bremen, Germany
[2] Department of Physics & Astronomy, University of Southampton, Southampton SO17 1BJ

July 16, 2002

**Abstract**

The hot plasma permeating clusters of galaxies often shows a central peak in the X-ray surface brightness that is coincident with a drop in entropy. This is taken as evidence for a cooling flow where the radiative cooling in the central regions of a cluster causes a slow subsonic inflow (e.g. Fabian, 1994). Searches in all wavebands have revealed significantly less cool gas than predicted (Böhringer et al., 2002) indicating that the mass deposition rate of cooling flows is much lower than expected. However, most cooling flow clusters host an Active Galactic Nucleus (AGN) at their centres (Burns, 1990). AGN can inflate large bubbles of hot plasma that subsequently rise through the cluster atmosphere, thus stirring this gas (Churazov et al., 2000, 2001). Here we report on the results from highly resolved hydrodynamic simulations which for the first time show that buoyant bubbles increase the cooling time in the inner cluster regions and thereby significantly reduce the deposition of cold gas. This work demonstrates that the action of AGN in the centres of cooling flow clusters can explain the observed absence of large quantities of cooled gas.

The explanation of the absence of cold gas in the gaseous atmospheres of cooling flow clusters strongly suggests the presence of an, at least periodically active, source of heat in their centres. Heating the gas at the cluster centre increases its entropy and thus its radiative cooling time. However, not all heating processes will explain the observations. Fully convective heating of the cluster gas would lead to the entropy being highest in the cluster centre, opposite to what is observed (Böhringer et al., 2002).

It has been suggested that the removal of cold gas and the feedback of energy through star formation and related supernovae can explain some of the observed properties of clusters (e.g. Voit & Bryan, 2001). While this is a promising scenario for the initial formation of clusters within the framework of hierarchical structure formation in the universe, the level of star formation found in detailed investigations of present-day cooling flow clusters is insufficient to explain the mass deposition rates inferred from X-ray observations (Böhringer et al., 2002). Moreover, the amount of star formation during the evolution of clusters necessary to explain the distribution of gas observed in them today, is difficult to reconcile with the observed abundances of heavy elements (Wu et al., 2000).

Radio-loud AGN drive strong outflows in the form of jets that inflate bubbles or lobes. The lobes are filled with hot plasma and can heat the cluster gas in various ways. In the case of very energetic jets the expansion of the lobes will be supersonic. The resulting strong shock will heat and compress the gas (Heinz et al., 1998; Kaiser & Alexander, 1999; Reynolds et al., 2001). However, there is mounting evidence that weaker jets, which are presumably much more common, do not lead to efficient shock heating (Fabian et al., 2000). Nevertheless, they still produce pockets of low-density gas that rise in the cluster atmosphere due to buoyancy. As the timescale for the rise of these bubbles is long compared to the lifetime of, both, the AGN and of the synchrotron emitting particles, we expect to observe only very few rising bubbles in the radio band (Brüggen & Kaiser, 2001). A substantial number of bubbles have been observed in a number of clusters as depressions in the X-ray surface brightness, sometimes associated with a low level of remnant radio emission (Heinz et al., 1998; McNamara et al., 2001). Moreover, it has been shown that a lot of energy in the form of bubbles may reside in clusters as the bubbles are difficult to detect (Brüggen et al., 2002). The buoyant rise and the subsequent mixing of high-entropy gas with the cluster material can lead to a significant re-structuring of the inner regions of a cooling flow. However, the details of the buoyant rise of the bubbles and the mixing with their gaseous environments are complex processes that are difficult to treat on the basis of analytical estimates alone.

Therefore, we performed highly resolved two-dimensional hydrodynamic simulations of buoyant gas in a typical cluster environment similar to previous simulations of lower resolution (Brüggen et al., 2002; Quilis et al., 2001). These simulations were performed using the FLASH code (Fryxell et al., 2000), which uses an explicit piece-wise parabolic method to solve the equations of compressible hydrodynamics. Only 2D simulations were performed because the primary aim was to achieve a very high spatial resolution in order to minimise the effects of numerical diffusivity that have hampered previous simulations. This was also aided by the use of an adaptive grid that places resolution elements only where they are needed such that an effective resolution of 4000 times 2000 elements could be achieved. Experience has shown that large-scale transport properties are not very different in 2D and 3D simulations so that we expect our conclusions to hold even for the three-dimensional reality.

The initial mass and temperature distributions were modeled on the well-studied Virgo Cluster (Nulsen & Böhringer, 1995) while the initial gas density profile was determined assuming hydrostatic equilibrium. Cosmological simulations as well as observations suggest that the gas in many clusters may not be in an equilibrium state. However, gas flows within a cluster of low Mach numbers will be unable to impede the rise of the bubble and are more likely to increase than to decrease the mixing. On the other hand, clusters that are far from hydrostatic equilibrium are generally not expected to host cooling flows. To mimic the energy input by a jet, hot, buoyant gas was injected continuously into a spherical region off-set from the gravitational centre with zero initial velocity. Fig. 1 shows a logarithmic contour plot of the density at 60 and 120 Myrs after the start of the energy injection for run 2. We have also performed a run (run 1) with a lower rate of energy injection. In both cases, one can see how the buoyant plume rises upwards while Kelvin-Helmholtz and secondary Rayleigh-Taylor instabilities form tori and eddies thus mixing the bubble gas with the colder cluster gas.

The horizontally averaged entropy as a function of time is shown in Fig. 2. Note that the entropy at lower radii is increased at the expense of the entropy at larger radii. As all motions at this stage are subsonic and there are no shocks, the increased entropy at the cluster centre comes entirely from mixing. Shortly after the bubble has crossed a particular layer, the entropy starts to decrease because lower entropy gas from closer to the cluster centre is being uplifted in its wake. Then the entropy rises again because the lower entropy material continues to be uplifted and higher entropy material is being funneled down to make place for the uplifted gas. This leads to a net increase in entropy in the lower regions. The increase in entropy in the bottom 20 kpc of the cluster is about 17.5run 2 over the course of the simulation. This implies that mixing is almost as effective in run 1 as in run 2 even though 13 times as much energy is injected in run 2. In both runs the mixing process is gentle enough not to reverse the entropy gradient. In run 2 the entropy in the bottom layers starts to decrease again after the jet has been switched off and low entropy material starts to flow back to the centre to re-establish hydrostatic equilibrium.

Fig. 3 shows the changes of the radiative cooling time of the cluster gas in the horizontal slices. Initially, the cooling time in the outer regions decreases as the rising bubble compresses the gas in front of it. However, the cooling time in the central regions increases as high-entropy material from larger radii is being mixed into the centre. Meanwhile, the low-entropy gas lifted from the cluster centre expands adiabatically as a result of which its cooling time increases as well. In run 2 the cooling time in the bottom slice went up by as much as 40 original cluster gas alone. When the gas in the hot bubbles is taken into account as well, the cooling time increases even more. In the cooling flow model, mass deposition is greatest in the core where the cooling time drops sharply. The mass that condenses out in the cores ceases to provide pressure support and thus causes a slow inflow. Therefore, the cooling flow can be delayed significantly if the cooling of gas in the central regions can be halted.

We have found that the average temperature in the core goes up as hotter material from outer regions is mixed into core. For the parameters chosen here, the rate of increase in temperature is very close to the cooling rate by bremsstrahlung. To be more precise, we have computed the ratio between the heating time , where the temperature T and its time derivative are averaged over concentric annuli about the centre, and the radiative cooling time. It was found that within the central 20 kpc this ratio varied between 1.02 1.15 which means that the heating is able to match the radiative cooling very closely. This leads us to conclude that, both, the decrease of the cooling time and the mixing of hotter material into the core causes a significant decrease in the mass deposition rate compared to the rate predicted by the quasi-static cooling flow model.

Clearly, the lifetime of the central AGN activity is short compared to the evolutionary timescale of the cluster gas. Therefore, once the AGN stops supplying energy to the buoyant bubbles, the cluster gas will settle down once again and a full cooling flow may be re-established. The cooling gas flowing to the very centre of the cluster may then trigger a further active phase of the AGN. Thus a self-regulating process with cooling periods alternating with brief bursts of AGN activity may be established (Binney & Tabor, 1995; Soker et al., 2001). It has been found that 71clusters show evidence for radio activity (Burns, 1990). If all AGN at the centres of cooling flows go through activity cycles of constant length, ton, separated by periods of quiescence, again of constant length, toff, then these observations imply that toff 0.41 ton. Figures 2b and 3b indicate that the simulated cluster (run 2, ton=239 Myrs) starts to relax back to its original state and at time ton+ toff =337 Myrs may well get close to a similar configuration when the AGN activity started. Therefore, the cluster gas may never evolve beyond a state conducive to the start of AGN activity. This may explain the absence of gas in clusters with a temperature below 1 keV. The high effective resolution and the accuracy of the numerical scheme used here have made it possible for the first time to make firm conclusions about the heating and mixing of the intra-cluster medium.

## Acknowledgments

The computations reported here were performed using the UK Astrophysical Fluids Facility (UKAFF). The software used in this work was in part developed by the DOE supported ASCI/Alliances Center for Thermonuclear Flashes at the University of Chicago.

## References

Binney J., Tabor G., 1995, MNRAS, 276, 663

Böhringer H., Matsushita K., Churazov E., Ikebe Y., Chen Y., 2002, A&A, 382, 804

Brüggen M., Kaiser C. R., 2001, MNRAS, 325, 676

Brüggen M., Kaiser C. R., Churazov E., Enßlin T. A., 2002, MNRAS, 331, 545

Burns J. O., 1990, AJ, 99, 14

Churazov E., Brüggen M., Kaiser C. R., Böhringer H., Forman W., 2001, ApJ, 554, 261

Churazov E., Forman W., Jones C., Böhringer H., 2000, A&A, 356, 788

Fabian A. C., 1994, ARA&A, 32, 277

Fabian A. C., Sanders J. S., Ettori S., Taylor G. B., Allen S. W., Crawford C. S., Iwasawa K., Johnstone R. M., Ogle P. M., 2000, MNRAS, 318, L65

Fryxell B., Olson K., Ricker P., Timmes F. X., Zingale M., Lamb D. Q., MacNeice P., Rosner R., Truran J. W., Tufo H., 2000, ApJ Supp., 131, 273

Heinz S., Reynolds C. S., Begelman M. C., 1998, ApJ, 501, 126

Kaiser C. R., Alexander P., 1999, MNRAS, 305, 707

McNamara B. R., Wise M. W., Nulsen P. E. J., David L. P., Carilli C. L., Sarazin C. L., O'Dea C. P., Houc J., Donahue M., Baum S., Voit M., O'Connell R. W., Koekemoer A., 2001, ApJ, 562, L149

Nulsen P. E. J., Böhringer H., 1995, MNRAS, 274, 1093

Quilis V., Bower R. G., Balogh M. L., 2001, MNRAS, 328, 1091

Reynolds C. S., Heinz S., Begelman M. C., 2001, ApJ, 549, L179

Soker N., White R. E., David L. P., McNamara B. R., 2001, ApJ, 549, 832

Voit G. M., Bryan G. L., 2001, Nat., 414, 425

Wu K. K. S., Fabian A. C., Nulsen P. E. J., 2000, MNRAS, 318, 889

# Figures

**Figure 1:** Snapshots of the density at two different times in our simulation (run 2). The left contour plot shows the density after 60 Myrs and the right one after 120 Myrs. The scale is logarithmic and unit density corresponds to $3.4 \times 10^{24}$ g cm$^{-3}$. The computational domain covered a rectangular region of 50 kpc in width and 100 kpc in height. The injection region had a radius of 1 kpc and the rate of energy input was L= $4.1 \times 10^{43}$ erg s−1. The energy input was switched off after 229 Myrs. The temperature of the injected fluid was 100 times greater than the temperature of the ambient fluid at the same location. All simulations were run on 27 processors on an SGI Origin 3000. We have also performed a run (run 1) with a lower rate of energy injection (L= $2.1 \times 10^{42}$ erg s$^{-1}$) where the energy input was switched off after 334 Myr. At all boundaries outflow conditions were chosen. The gravitational potential was taken as static in order to save solving Poisson's equation at each time step. The gas was treated as a single fluid with a polytropic equation of state and magnetic fields have been ignored. Radiative cooling was neglected as the cooling time in the central regions of the cluster is still longer than the total simulated time. Numerical experiments have shown that the neglect of cooling does not influence the large-scale dynamics of the fluid.

**Figure 2:** Average entropy of the ambient material at different heights. Material with entropy below a suitably chosen entropy threshold can be identified as initially ambient material which enables us to study the temperature and entropy evolution of the original cluster gas alone. For this purpose we divided the computational domain into 10 horizontal slices of 10 kpc thickness and computed averages of physical quantities in each slice. Each curve corresponds to a separate slice which is labelled with the height of its lower boundary. The left panel corresponds to run 1 and the right one to run 2.

**Figure 3:** Cooling time of the ambient material averaged over horizontal slices. Each curve corresponds to a separate slice which is labelled with the height of its lower boundary. For each slice the cooling time is normalised to its value at the start of the simulation. The left panel corresponds to run 1 and the right one to run 2.

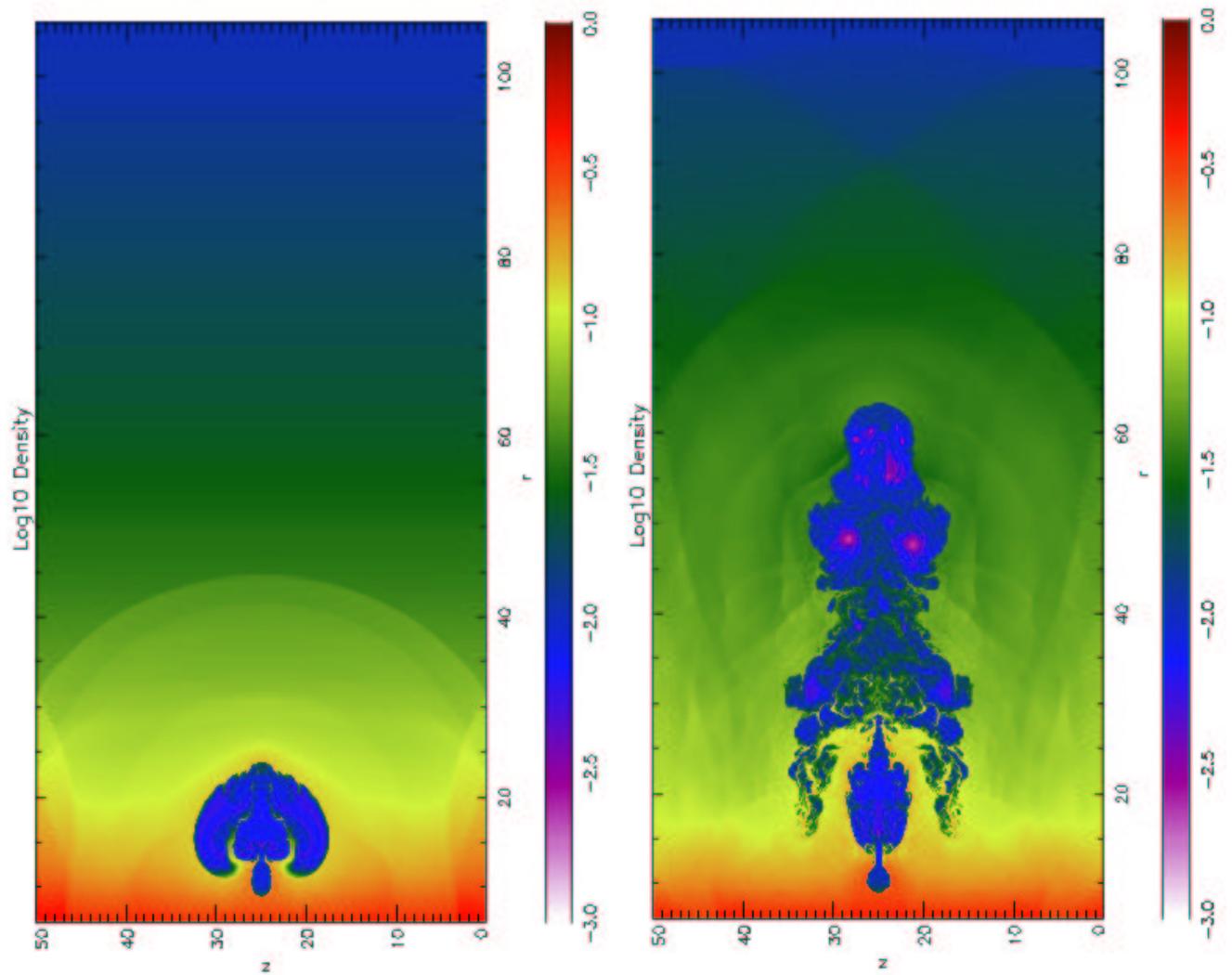

Figure 1

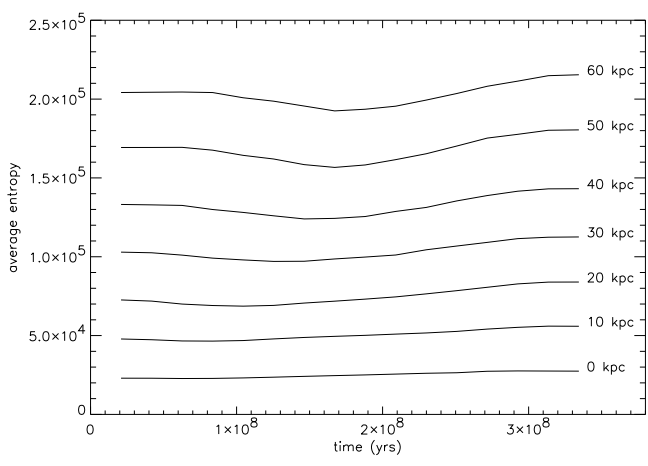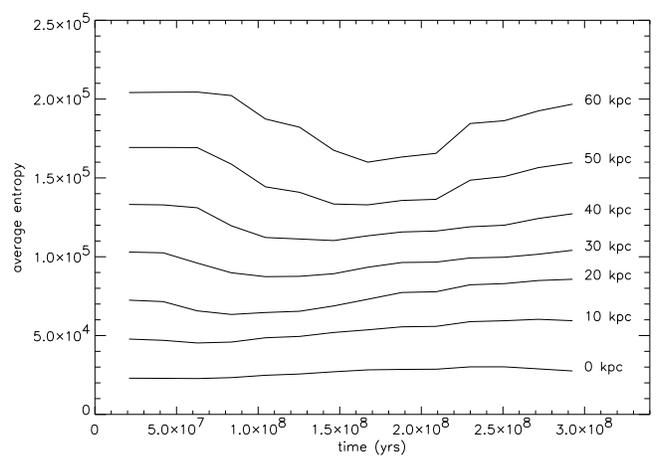

**Figure 2**

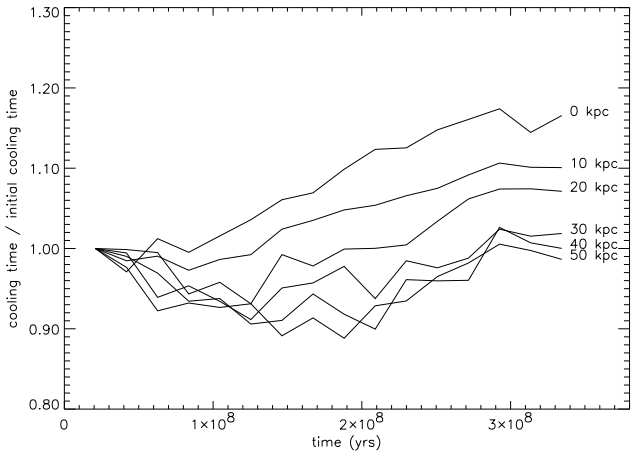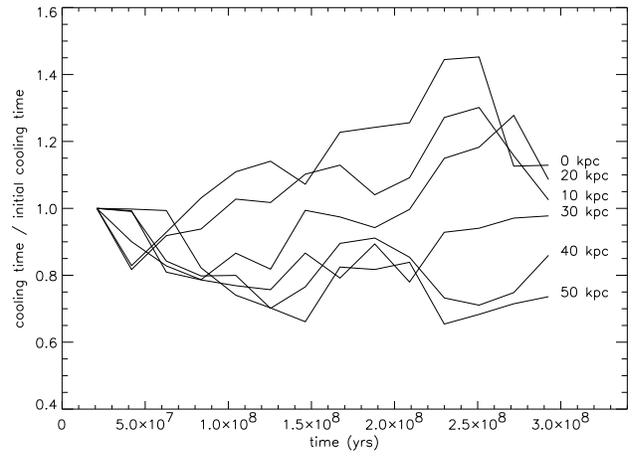

**Figure 3**

5